# Panoramic-reconstruction temporal imaging for seamless measurements of slowly-evolved femtosecond pulse dynamics

Bowen Li,[1,2,†] Shu-Wei Huang,[1,*,†] Yongnan Li,[1,3] Chee Wei Wong,[1,*] and Kenneth K. Y. Wong[2,*]

[1] Fang Lu Mesoscopic Optics and Quantum Electronics Laboratory, University of California, Los Angeles, CA 90095 USA

[2] Department of Electrical and Electronic Engineering, The University of Hong Kong, Pokfulam Road, Hong Kong, China

[3] School of Physics and The MOE Key Laboratory of Weak Light Nonlinear Photonics, Nankai University, Tianjin, China.

[†] These authors contributed equally to this work

[*] Email: swhuang@seas.ucla.edu; cheewei.wong@ucla.edu; kywong@eee.hku.hk

Single-shot real-time characterization of optical waveforms with sub-picosecond resolution is essential for investigating various ultrafast optical dynamics. However, the finite temporal recording length of current techniques hinders comprehensive understanding of many intriguing ultrafast optical phenomena that evolve over a time scale much longer than their fine temporal details. Inspired by the space-time duality and by stitching of multiple microscopic images to achieve a larger field of view in the spatial domain, here a panoramic-reconstruction temporal imaging (PARTI) system is devised to scale up the temporal recording length without sacrificing the resolution. As a proof-of-concept demonstration, the PARTI system is applied to study the dynamic waveforms of slowly-evolved dissipative Kerr solitons in an ultrahigh-$Q$ microresonator. Two 1.5-ns-long comprehensive evolution portraits are reconstructed with 740-fs resolution and dissipative Kerr soliton transition dynamics, in which a multiplet soliton state evolves into stable singlet soliton state, are depicted.

The capability of characterizing arbitrary and non-repetitive optical waveforms with sub-picosecond resolution at single-shot and in real-time is beneficial for different fields such as advanced optical communication [1, 2], ultrashort pulse generation [3, 4], optical device evaluation [5] and ultrafast bio-imaging [6-8]. Moreover, it has helped to unveil fascinating ultrafast phenomena in optics, such as the onset of mode-locking [9, 10], soliton explosions [11-12] and optical rogue waves [13-15], as well as many other fields [16-18]. Temporal imaging is one of the most promising techniques perceived and developed to meet the need of single-shot real-time waveform characterization [7, 8, 14, 15, 19-28]. Based on space-time duality [19-21], quadratic phase modulation (time lens) and dispersion can be properly combined to significantly enhance the temporal resolution [14, 15, 22-26] and a record value of 220 fs has been demonstrated [26]. On the other hand, just like there is always a limitation on the field-of-view in any spatial imaging systems, the single-shot recording length of temporal imaging systems has been hitherto limited to less than 300 ps [23]. Owing to this limitation, the time-bandwidth product (TBWP, the ratio between the recording length and the temporal resolution) of the state-of-the-art temporal imaging systems has not exceeded 450 [26]. Such situation hinders the applications of temporal imaging systems to study many important optical nonlinear dynamics where not only fine temporal details but also long evolution information are necessary for a comprehensive understanding of the phenomena. For example, studying the dynamics of dissipative Kerr solitons [29-31] is of particular interest because of their potential applications in low-phase noise photonic oscillators [32, 33], broadband optical frequency synthesizers [34, 35], miniaturized optical clockwork [36], and coherent terabit communications [37]. While the soliton generation benefits greatly from the ultrahigh quality factor ($Q$) of the microresonator, the ultrahigh $Q$ also renders its formation and transition dynamics slowly evolved at a time scale much longer than the cavity roundtrip time [38, 39], which causes significant challenges in the experimental real-time observation. Similarly, an optical metrology system that combines the feats of fine temporal resolution and long measurement window is also desired in the study of optical turbulence and laminar-turbulent transition in fibre lasers [40, 41], which leads to a better understanding of coherence breakdown in lasers and laser operation in far-from-equilibrium regimes. To capture comprehensive portraits of these processes, as well as many other transient phenomena in nonlinear optical dynamics [14, 15, 42, 43], a temporal imaging system with a TBWP much greater than 1,000 is necessary.

While the most straightforward way to implement a time lens is to use a phase modulator, the TBWP using this approach is fundamentally limited by the maximum achievable modulation depth and is typically less than 10 [44-46]. Alternatively, a time lens can be constructed all-optically through cross-phase modulation [47, 48]. However, similar limitations exist, since large modulation depth requires high pump power, which in turn induces self-phase modulation (SPM) on the pump pulse and distorts the temporal intensity envelope. Consequently, the reported TBWPs using this approach are only around 20 [47, 48]. Therefore, state-of-the-art temporal imaging systems are mostly implemented through parametric mixing with a linearly-chirped pump pulse, where TBWPs up to several hundred have been achieved [7, 8, 14, 15, 22-28]. The practical limitation on further improvement of the TBWP in parametric temporal imaging systems originates from the maximum effective pump bandwidth and the maximum pump dispersion [20, 21]. While the effective pump bandwidth is restricted by phase-matching condition in parametric conversion, excessive pump dispersion degrades system performance by inducing both large third-order-dispersion (TOD) aberration and undesired propagation loss. Therefore, it is impractical to substantially improve the TBWP of temporal imaging systems under conventional configuration. Meanwhile, limitations on TBWP also exist for other techniques that achieve comparable performance [49-53]. Single-shot real-time spectral interferometry [53] has been adopted to reconstruct the time-domain information, achieving a temporal resolution of about 400 fs. However, its temporal recording length is limited by the spectral resolution (10 pm) to around 350 ps, which results in a TBWP of 875. Another measurement technique combines spectral slicing of the optical signal with parallel optical homodyne detection using a frequency comb as a reference [52]. Even though a TBWP larger than 320,000 has been demonstrated at a temporal resolution of about 6-ps, it is practically challenging to reach the sub-picosecond regime. Acknowledging current existing methods, a waveform measurement technique achieving sub-picosecond temporal resolution and long temporal recording length is urgently needed and it will be a powerful tool for studying ultrafast dynamics in different areas.

In order to achieve this goal, we propose and experimentally demonstrate a panoramic-reconstruction temporal imaging (PARTI) system, in analogy with the wisdom of stitching multiple mosaic images to achieve larger-field-of-view in the spatial domain [54, 55]. The PARTI system consists of a high-fidelity optical buffer, a low-aberration time magnifier and synchronization-control electronics. Through the PARTI system, different parts of a transient optical dynamic waveform can be characterized sequentially in multiple steps. After signal processing, a magnified panoramic image of the original waveform is reconstructed from multiple mosaic images. A temporal recording length of 1.5 ns is realized without sacrificing the 740-fs resolution, thus achieving a TBWP of over 2,000, about 5 times larger than the record value previously demonstrated in conventional temporal imaging systems [26]. As a proof-of-concept demonstration, the PARTI system is applied to observe the dissipative Kerr soliton transition dynamics in an ultrahigh-$Q$ microresonator and two distinct multiplet-to-singlet dissipative Kerr soliton transition dynamics are observed.

# RESULTS

**Principle of operation**

Figure 1 shows a simulated example of dissipative Kerr soliton dynamics and describes how the PARTI system captures the slowly-evolved process in a single-shot manner. As shown in Figure 1a, in the governing Lugiato-Lefever formalism [56, 57], the dissipative Kerr soliton dynamics is depicted in a two-dimensional (2D) space spanning by the cavity time $\tau$ and the evolution time $t$. While the temporal structure of the intra-cavity field is detailed in the $\tau$ dimension at the sub-picosecond time scale, the evolution and transition dynamics is portrayed in the $t$ dimension at a much longer nanosecond time scale, which is associated with the cavity photon time of the microresonator. At the beginning of the evolution, the cavity exhibits a triplet soliton state. However, at around 1 ns, the top two solitons start to be attracted to each other and finally merge into a single soliton at around 1.2 ns. The bottom soliton also shifts upwards during the soliton fusion. After 1.6 ns, a stable doublet soliton state is reached, where both solitons exhibit higher intensity owing to the energy conversion inside the microresonator. To comprehensively characterize this soliton-fusion process, a recording length of at least 1 ns is desired, while a sub-picosecond temporal resolution is required to effectively resolve the soliton shape. Therefore, a TBWP larger than 1000 is necessary.

Figure 1b shows how the PARTI system overcomes the limitation of TBWP in conventional temporal imaging systems and thus captures the slowly-evolved soliton dynamics. The signal under test (SUT) is a pulse train that schematically represents the 2D evolution in Figure 1a. Since the SUT is transient and non-repetitive, the concept of sample scanning in the spatial domain cannot be conveniently adopted in temporal imaging systems. To address this problem, a fibre-loop based optical buffer is integrated with a time magnifier to realize temporal scanning using stroboscopic signal acquisition [58, 59], a technique commonly adopted in sampling oscilloscopes. As shown in Figure 1b, the optical buffer creates multiple identical replicas of SUT with a constant time interval, which will be subsequently measured by the following time magnifier, thus realizing the temporal scanning on a transient SUT. Using the optical buffer, SUT replicas can be generated with a pre-defined period of $T_1$. If the measurement period of time magnifier is $T_2$, then in each frame, the time magnifier captures a different section of the long waveform with a step size equal to $|T_1-T_2|$. Furthermore, by matching the step size to the recording length of the time magnifier, seamless measurement of a long waveform can be realized. The output of the PARTI system represents the magnified waveform corresponding to different sections of the long SUT and is recorded by a high-speed real-time oscilloscope. After data processing, neighbouring frames of magnified waveform will be stitched together to reconstruct a magnified panoramic image of the original SUT. Therefore, the effective single-shot recording length is scaled by the number of replicas without sacrificing the temporal resolution, thus substantially enhancing the TBWP.

**Low-aberration time magnifier**

The foundation to construct the PARTI system is a parametric time magnifier with low aberration. The four-wave mixing (FWM) process was chosen as opposed to other parametric processes because it allows high-quality processing of SUT, pump and output simultaneously in the telecommunication band [21]. In addition, since multiple frames of magnified waveform need to be stitched together to obtain the panoramic image, it is critical to ensure a stable impulse response across the recording window of the time magnifier, *i.e.* a low-aberration FWM time magnifier. For an in-focus time magnifier, the main aberration comes from the TOD in the dispersive path for input and pump [60]. In order to construct a low-aberration time magnifier with long recording length, the combination of dispersion compensating fibre (DCF) and large effective-area fibre (LEAF) is used to achieve large linear dispersion (fourth and higher order dispersion neglected). As shown in the experimental setup in Figure 2a, both the input dispersion and the pump dispersion is provided by combining DCF and LEAF. Since the LEAF has the opposite dispersion slope [0.08 ps·nm$^{-2}$km$^{-1}$] compared to the DCF [-0.598 ps·nm$^{-2}$km$^{-1}$], combining the two types of fibre according to the ratio of their dispersion slope results in linear net dispersion. Moreover, LEAF features in very small dispersion-to-dispersion-slope ratio ($K_{LEAF}$=D/S=45 nm) compared to standard single-mode fibre (SMF) ($K_{SMF}$=D/S=275 nm). Therefore, using a LEAF fibre to compensate dispersion slope of DCF sacrifices much less net dispersion compared with using SMF, which facilitates achieving large linear dispersion with moderate insertion loss. In the current system, the SUT is dispersed for 35 ps$^2$ before being combined with the pump through the wavelength-division multiplexer (WDM). In the lower branch of the system, a broad-band mode-locked laser (MLL) goes through a dispersion of 71.2 ps$^2$ and is then pre-amplified by a low-noise EDFA. The following band-pass filter selects the spectral component from 1555 nm to 1565 nm, which is subsequently amplified again to 100 mW to generate the pump for the time magnifier. The pump and SUT are launched together into the highly-nonlinear fibre (HNLF), and the generated idler is filtered out and goes through the output dispersion (2152.5 ps$^2$), which is then amplified again to become the final output of the time magnifier. Overall, the system satisfies the imaging condition [20]

$$\frac{-1}{\Phi_1^{''}} + \frac{1}{\Phi_2^{''}} = \frac{1}{\Phi_f^{''}} \tag{1}$$

where the $\Phi_1^{''}$ (35 ps$^2$), $\Phi_2^{''}$ (2152.5 ps$^2$), and $\Phi_f^{''}$ (35.6 ps$^2$) are the input, output, and focal group-delay dispersions, respectively while the minus sign originates from the phase conjugation during the chosen parametric process. Therefore, the temporal magnification ratio

is

$$M = \frac{\Phi_2^{''}}{\Phi_1^{''}} = 61.5 \qquad (2)$$

To characterize the performance of the time magnifier, a femtosecond pulse with 10-nm bandwidth and centre wavelength of 1543 nm is used as input of the system. As shown in Figure 2b, in the FWM spectrum after the HNLF, a narrow-band idler is generated, which is then filtered out and became the temporally magnified signal after output dispersion. The femtosecond pulse is shifted temporally across 300 ps input window (input time scanning) and the corresponding output waveform is recorded. As shown in Figure 2c, the output time is linearly proportional to the input time with a slope of 61.5. Moreover, the output pulse width during input time scanning is stable at 54 ps. The corresponding output pulse shape (intensity normalized individually) is also almost identical (Figure 2d), which indicates very small aberration from TOD. This feature is most critical for implementing the temporal scanning microscope, as ideally the overlapping areas should be identical in neighbouring measurement frames so as to be clearly identified for image stitching. The small fluctuating tail of the waveform results from the impulse response of the photodetector, which is shown in the inset. Based on the results above, the average measured pulse width is 54 ps/61.5 = 878 fs. Since the real pulse width of the input signal is measured to be 470 fs through auto-correlation, the de-convolved impulse response or the temporal resolution of the time magnifier is calculated to be $\sqrt{878^2 - 470^2} = 740$ fs. Therefore, the low-aberration time magnifier achieves a large time-bandwidth product of 300 ps/740 fs = 405.4 enabled by the large linear dispersion links in the system, which is comparable to the largest time-bandwidth product previously demonstrated in temporal imaging systems [26].

**Optical buffer and timed replication**

In order to generate multiple replicas of SUT for subsequent stroboscopic signal acquisition (discussed in next section), a fibre-loop based optical buffer is designed and the experimental setup is shown in Figure 3a. During operation, a section of waveform will be carved out by amplitude modulator 1 (AM1) and loaded into the buffer through a 50/50 coupler. After each circulation inside the fibre-loop cavity, 50% of the buffered waveform is coupled out as a replica, while the other 50% is circulated for the next round. The total cavity length is designed to be around 8.2 m and the cavity period can be fine-tuned from 39.7 ns to 40 ns using the optical delay-line in order to match the frame rate of the time magnifier. AM2 functions as a switch by controlling the intra-cavity loss. The switch is turned on only when the SUT passes the AM2 and therefore, the AM2 controls the number of replicas generated from the buffer. More importantly, the periodic switching of AM2 prevents the self-lasing operation of the optical buffer, which substantially suppresses the amplification noise during the buffering. Additionally, a WDM filter with a passband from 1537 nm to 1547 nm further minimizes the buffering noise. A 2-m erbium-doped fibre (EDF) pumped by 980-nm laser diode provides a maximum gain of around 20 dB to compensate the total cavity loss ($\approx$ 12 dB). To minimize the dispersion distortion, 0.5-m DCF is added to the cavity and the net dispersion of the buffer is measured to be around $6.12 \times 10^{-3}$ ps$^2$ (see Supplementary Note 1 for details), which corresponds to the dispersion of only 0.28-m SMF. For a 740-fs optical pulse (equal to the resolution of the time magnifier), such residual dispersion will only result in less than 5% pulse shape distortion after 10 roundtrips. Therefore, the influence of residual net dispersion is small enough to be neglected. Finally, by optimizing the polarization controllers (PC) both outside and inside the cavity, the buffer generates high-fidelity replicas of the input waveform.

To visualize the performance of the buffering, arbitrary waveforms (See Supplementary Note 2 for details) generated from an ultrahigh-$Q$ microresonator are used as SUT and launched into the optical buffer to generate 10 replicas. All the SUT have the duration of ≈ 5 ns and spectral bandwidth of ≈ 10 nm, but the waveforms shapes are distinct from each other. Figure 3b shows the output waveform of ten replicas for a certain SUT after the buffering. The shape as well as the intensity of the SUT are well preserved during buffering. In Figure 3c, the ten replicas in Figure 3b are overlapping together (grey curves) and are compared to the averaged waveform (blue curve). It is obvious that the optical buffer can generate high fidelity replicas, which only exhibit small fluctuations (less than 10%) during each buffering compared to the averaged reference. Figure 3d and 3e show similar performance for two more arbitrary examples of SUT. To quantitatively evaluate the buffering fidelity, a total of 20 different SUTs are tested. As shown in Figure 3f, for each SUT, the cross-correlation coefficient between different replicas and the first replica (original waveform) are calculated and represented by purple columns, while the blue triangles shows the average value of the 20 SUTs. The first column set represents the cross-correlation coefficient of the first replica with itself (*i.e.* auto-correlation) and therefore the value equals 1 for all 20 SUTs. After the first buffering time, the coefficients start to decrease gradually with each buffering, which indicates larger and larger deviations from the original waveforms owing to the buffering distortions. However, even for the tenth replica, the average cross-correlation coefficient is still larger than 0.99. Therefore, the optical buffer is able to generate high-fidelity replicas for arbitrary temporal waveforms. As the dispersion broadening is far below the temporal resolution of direct measurement (≈18-GHz bandwidth), the majority of the deviation is attributed to the amplification of noise and gain narrowing effect in the buffer and thus reducing the cavity loss and inclusion of gain equalizers can further improve the quality of the optical buffer, if necessary.

**Stroboscopic signal acquisition and the PARTI system**

Equipped with the low-aberration time magnifier and high-fidelity optical buffer, the PARTI

system is implement based on the concept of stroboscopic signal acquisition [58, 59]. The basic idea has already been illustrated in Figure 1. By inducing a period difference between the buffered replicas and the pump pulses, the time magnifier captures a different section of SUT consecutively on each replica. In this way, a long SUT can be fully scanned in multiple steps and the complete waveform can be reconstructed from the magnified waveform of each stroboscopic acquisition. The experimental detail of implementation is shown in Figure 4a. In order to emphasize the key components for stroboscopic acquisition, the synchronization electronics are highlighted while the optical buffer and the time magnifier are simplified and slightly shadowed. The key electronics can be divided into the following three groups. First of all, a repetition-rate-stabilized femtosecond fibre MLL and a 1.2-GHz photodetector together generate a 250-MHz electrical clock signal, which serves as the time base of the whole system. Secondly, an arbitrary waveform generator (AWG), and a delay generator create electrical patterns that control the stroboscopic acquisition. Finally, the three AMs convert the electrical patterns to the optical domain, which control the SUT loading (AM1), optical-buffer switching (AM2) and time-magnifier-pump generation (AM3) respectively.

The detailed timing chart of the system is shown in Figure 4b. As indicated by the vertical blue dashed line, the whole system is operated with a frame rate of 2 MHz. In every 500 ns, the AM1 will load from input a 5-ns-long waveform as SUT (first horizontal axis). After the SUT is loaded into the buffer, AM2 will be switched on only when the SUT arrives in each circulation. Therefore, in the second horizontal axis, AM2 opens every 40 ns and generates ten identical replicas in each 500-ns frame. Ideally, the separation of each gating should be identical with the cavity period of the buffer (39.85 ns). But limited by the sampling speed of AWG (1 Gs·s$^{-1}$), the separation is set as 40 ns. However, since each SUT is only circulated for 10 times inside the buffer and the gating width (10 ns) is much broader than the SUT duration, the slight mismatch between the gating period and the cavity period will not influence the performance of the buffer. After the buffering, ten replicas will be generated with a separation equal to the cavity period (fourth horizontal axis). AM3 performs pulse-picking on the MLL to generate a pump for the time magnifier every 40 ns (third horizontal axis). The corresponding real electrical driving signals for three AMs are shown in Figure 4c. Owing to the period difference (150 ps) between the time magnifier and the SUT replicas, the time magnifier will scan the SUT from left to right with a step of 150 ps, thus realizing the stroboscopic signal acquisition.

To directly visualize the stroboscopic signal acquisition, amplified spontaneous emission (ASE) from an EDFA is used as SUT and combined with time-magnifier pump when the whole system is operated according to the timing chart. The corresponding waveform is shown in Figure 4d. As observed in the left inset, in each 500-ns period, the first replica of the waveform section (broad and flat pedestal, orange) is aligned with the pump (sharp peak, blue) on the left side while in the last frame, the time-magnifier pump is already scanned to the right side, as shown in the right inset. In this way, the ten output frames will be generated in each 500-ns period, which corresponds to the magnified waveform at 10 consecutive positions of the SUT. By identifying the overlapping areas of the output waveform in neighbouring frames, the 10 sections of magnified waveform can be stitched together to reconstruct a much longer continuous waveform (see Supplementary Note 3 for details). Consequently, the recording length of the time magnifier can be scaled by the number of replicas while maintaining the high temporal resolution. Overall, the current PARTI system demonstrates a TBWP of more than 2000, about 5 times larger than the recording value achieved to date in temporal imaging systems [26].

**Measurement of dissipative Kerr soliton dynamics**

Finally, to demonstrate the capabilities of the PARTI system, the system is applied to observe the dynamic evolution of dissipative Kerr solitons inside an ultrahigh-$Q$ microresonator. The corresponding FWM spectrum measured after the time lens is shown in Supplementary Figure 3. The final output of the system is detected by an 18-GHz photodetector and then digitized and recorded by a real-time oscilloscope. After data processing (see Supplementary Note 3 for details) on the measurement results, two sections of 1.5-ns-long waveform with a 740-fs resolution are reconstructed, which represent a TBWP of more than 2000. With the unprecedented measurement capability, fascinating dissipative Kerr soliton dynamics in a high-$Q$ microresonator is observed. To clearly visualize the evolution details, we section the one-dimensional waveform according to the cavity roundtrip time (11.29 ps) of the microresonator to rearrange the data into a two-dimensional matrix and create 2D evolution portraits to depict the dissipative Kerr soliton transition dynamics.

In the first case, a transition process that resembles the simulation result in Fig. 1(a) is observed. As shown in Figure 5a, at the beginning stage (0 ps to around 400 ps), three solitons (triplet state) with almost equal intensity exist in the cavity. Figure 5b plots the waveforms at three different time slices of 0 ps (black), 113 ps (blue) and 237 ps (red), which shows that the triplet solitons roughly maintain their intensities and positions in the cavity throughout the beginning stage. The three curves were vertically offset for clarity and vertical black dashed lines are plotted according to the soliton positions at 0 ps (black curve) to emphasize the position change of solitons at different time slices. After that, in the middle stage (400 ps to around 800 ps), the first two solitons starts to be attracted to each other and eventually merge into a singlet soliton at around 800 ps. The third soliton is also shifted upwards during the merging of the other two solitons, just like the simulation in Figure 1a. However, the third soliton does not survive during the transition and starts to fade after 500 ps. The soliton fusion details are shown in Figure 5c, where waveforms at 440 ps, 565 ps and 677 ps are shown. At these three specific time positions, the separation between the first two solitons evolves from 3.8 ps to 3 ps and then to 1.5 ps. After this transitioning middle stage, a singlet soliton state is achieved inside the cavity, and the state remains for more than 600 ps, or 53 cavity roundtrips. Similarly, three waveforms in this stage are shown in Figure 5d, indicating the high stability during the final stage. Black dashed curves emphasizing the soliton transition traces are plotted against the 2D portrait in Figure 5a and 5e, which is obtained by polynomial fitting the peak positions of the solitons.

In addition to the first example, a different dynamic process is also observed which also generates the singlet soliton state eventually but without soliton fusion. As shown in Figure 5e, in the first stage (0 to around 370 ps) two solitons co-exist in the cavity. In the meantime, the doublet solitons repulse each other slightly and the first soliton gradually fades away. At around 370 ps, the upper soliton disappears, but at the same time two other solitons emerge. In the second stage (370 ps to 1 ns), in contrast to the first stage, the triplet solitons are attracted to the centre slowly. At the end of the second stage, both the top and bottom solitons fade away, while the middle one survives and evolves into a singlet soliton with higher intensity in the final stage (1 ns to 1.5 ns). Similar to the first example, the singlet soliton state is much more stable compared to previous states and lasts over 500 ns. Again, waveforms at three different time slices are plotted together for each distinct stage, which shows weak pulse repulsion (f), weak pulse attraction (g) and stable single soliton state (h).

Notably, these two soliton dynamics are observed with the same excitation protocol in the same microresonator. It has been shown, both theoretically and experimentally [29, 39], that the dissipative Kerr soliton formation is not a deterministic process, and soliton states with different orders can be accessed with a certain probability. The reason for such a stochastic behaviour is that these dissipative Kerr solitons can only be generated after the transition from chaos states which by nature are sensitive to initial conditions and noise processes. Future application of PARTI can unveil the probability function and depict the route out of chaos, leading to a better understanding of dissipative Kerr soliton formation. Specific excitation protocols to avoid the chaotic region have been proposed and theoretically studied [61], and it can be realized and verified experimentally in conjunction with PARTI. Furthermore, PARTI can be applied to study other fascinating nonlinear dynamics including Kerr frequency comb generation beyond Lugiato-Lefever equation [62] and novel mode-locking dynamics via Faraday instability [63].

**Discussion**

As the first proof-of concept demonstration, the stroboscopic signal acquisition (temporal scanning) is performed conservatively to ensure the accuracy of waveform reconstruction. It is worth noticing that while the single-shot recording length of the time magnifier is as large as 300 ps, the current temporal scanning adopts a step-size of only 150 ps. Therefore, in two consecutive scanning steps, about 50% of the measurement results are repetitive, which are used as the reference for waveform stitching. Despite this conservative configuration, 1.5-ns-long waveforms are reconstructed with 740-fs resolution, representing a record-high TBWP of 2,027. The TBWP of the system will be further substantially scaled by reducing the repetitive percentage in neighbouring steps and by increasing the number of buffering times. Currently, the step size is limited by the non-uniform responsivity across the recording window of the parametric time lens (see Supplementary Figure 3d). Because of the worse SNR near the boundaries of recording window, a small step size is adopted during experiment so that the reconstructed long waveforms only consist of the high-quality waveforms in the centre area of the recording window. An optical source with higher spectral flatness and nonlinear media with smaller TOD will contribute to a more uniform responsivity across the recording window of the time lens. Ideally, with a flat responsivity, the step-size can be identical with the recording length (*i.e.* no overlapping areas) to reconstruct continuous dynamic waveforms. Moreover, using a better designed optical buffer, the number of buffered replicas can also be substantially increased (See Supplementary Note 4 for detailed analysis). For example, some optical buffers have demonstrated generating more than 100 replicas with acceptable SNR degradation [64, 65]. Therefore, under the scenario of non-overlapping scanning and 100 times buffering, the PARTI system can theoretically capture a 30-ns-long non-repetitive dynamic waveform with 740-fs resolution (TBWP larger than $4\times10^4$), which will serve as a powerful tool for studying different kinds of ultrafast optical dynamics. Moreover, the generalized idea of waveform replication combined with single-shot acquisition is also applicable to other measuring techniques, such as the real-time spectral interferometry [53]. Therefore, our technique not only represents an advanced temporal imaging system, but may also stimulate more analogous innovations in the family of single-shot ultrafast measurement techniques.

In conclusion, a panoramic-reconstruction temporal imaging (PARTI) system is developed by integrating a fibre-loop based optical buffer with a low-aberration time magnifier. In analogy to a conventional microscope achieving larger field-of-view by scanning the sample and stitching microscopic images, our technique provides the possibility to observe ns-long dynamic waveforms while maintaining the sub-picosecond temporal resolution, thus overcoming the limitation of TBWP in conventional temporal imaging systems. As a proof-of-concept demonstration, the PARTI system is applied to observe the dissipative Kerr soliton transition dynamics in an ultrahigh-*Q* microresonator. By buffering the selected waveform ten times and measuring the waveform in ten steps, 1.5-ns long evolution processes are reconstructed with 740-fs resolution, which represents a TBWP of over 2,000, about 5 times larger than the record value demonstrated to date in conventional temporal imaging systems [26]. Moreover, the TBWP in our technique is scalable to even higher values by using larger number of replicas. With our technique, two distinct multiplet-to-singlet dissipative Kerr soliton transition dynamics are observed. The capability in observing such intriguing phenomena using long recording length and high resolution will not only facilitate the study of dissipative soliton dynamics, but also various ultrafast dynamic processes in other fields as well.

**Methods**

**Experimental setup:** The optical buffer consisted of a 4-port 50/50 coupler, an optical delay line, an amplitude modulator, a WDM filter, a polarization controller, a 980/1550 WDM coupler, 0.5-m dispersion compensating fibre (DCF38), and 2-m erbium-doped fibre (ER30-4/125). The net dispersion was estimated by measuring the pulse-broadening of a femtosecond source with an intensity auto-correlator after single-passing the open-loop buffer. The cavity period of the buffer was measured by buffering a picosecond pulse and measuring the separation of buffered replicas. The pump of the temporal magnification system was generated from a 250-MHz optical frequency-comb source (Menlo FC1500-250-WG), which was bandpass-filtered (1554-1563nm) and pulse-picked by AM3 to 25 MHz. The input dispersion consisted of 200-m DCF (LLMicroDK) and 1.486-km LEAF (Corning) while the pump dispersion consisted of 400-m DCF and 2.97-km LEAF of the same kind. The ratio between the DCF and LEAF was carefully designed to achieve near-zero TOD. The output dispersion was provided by combining two dispersion compensating modules (Lucent DCM) with a total dispersion of -1.689 ns/nm. The final optical signal was measured by an 18-GHz photodetector (EOT 3500f) and subsequently digitized by a 20 GHz real-time oscilloscope (Tektronix MSO 72004C) with 100-Gs·s$^{-1}$ sampling rate. The stroboscopic signal acquisition was controlled using a 1.2 GHz photodetector (DET01CFC), an arbitrary waveform generator (Tektronix AWG520), and a delay generator (SRS DG645).

**Si$_3$N$_4$ microresonator fabrication:** First a 5 μm thick oxide layer was deposited via plasma-enhanced chemical vapour deposition (PECVD) on p-type 8" silicon wafers to serve as the under-cladding oxide. Then low-pressure chemical vapour deposition (LPCVD) was used to deposit an 800 nm silicon nitride for the ring resonators, with a gas mixture of SiH$_2$Cl$_2$ and NH$_3$. The resulting silicon nitride layer was patterned by optimized DUV lithography and etched down to the buried oxide layer via optimized reactive ion dry etching. Next the silicon nitride ring resonators were over-cladded with a 3 μm thick oxide layer, deposited initially with LPCVD (500 nm) and then with PECVD (2500 nm). The device used in this study has a ring radius of 250 μm, a free spectral range of 88.6 GHz, and a loaded quality factor $Q$ of ≈ 1,000,000.

**Dissipative Kerr soliton generation and modelling:** The microresonator was pumped by a frequency-tunable continuous-wave laser with an on-chip power of 600 mW. For dissipative Kerr soliton generation, the laser frequency was scanned with a tuning speed of 2 THz·s$^{-1}$, via control of the piezoelectric transducer, across the cavity resonance from the blue side of the resonance. The soliton dynamics was modelled by numerically solving the mean-field Lugiato-Lefever equation with the symmetric split-step Fourier method and the classical Runge-Kutta method. The simulation started from vacuum noise and the temporal resolution was set to 5 fs.

**Data availability.** The data that support the plots within this paper and the findings of this study are available from the corresponding author on request.

**Acknowledgements:** The authors acknowledge discussions with Dr. Baicheng Yao and Dr. Xiaoming Wei. This material is based upon work supported by the Air Force Office of Scientific Research under award number FA9550-15-1-0081, the Office of Naval Research (N00014-14-1-0041), the National Science Foundation (14-38147 and 15-20952), Research Grants Council of the Hong Kong Special Administrative Region, China (Project Nos. HKU 17205215, HKU 17208414, and CityU T42-103/16-N), National Natural Science Foundation of China (N_HKU712/16), Innovation and Technology Fund (GHP/050/14GD) and University Development Fund of HKU.

**Author contributions:** B.L. and S.W.H. proposed the technique and designed the experiment. B.L., S.W.H., and Y.L. performed the experiment. B.L., and S.W.H. wrote the manuscript. All authors contributed to the analysis of the data and discussion and revision of the manuscript.

**Additional information:** The authors declare no competing financial interests. Correspondence and requests for materials should be addressed to S.W.H., C.W.W. and K.K.Y.W.

**Figure 1 | Working principle of the PARTI system. a,** Slowly-evolved dissipative Kerr soliton dynamics in an ultrahigh-$Q$ microresonator, obtained by numerically solving the Lugiato-Lefever equation. The orders-of-magnitude difference in the time scale between the cavity time and the evolution time poses an experimental challenge to capture the comprehensive picture of dynamics. **b,** The schematic of the panoramic-reconstruction temporal imaging (PARTI) system. The optical buffer generates multiple replicas (represented by blue, green and red, respectively) of the signal under test (SUT) and the subsequent time magnifier captures different portions of the SUT waveform on each replica. After data processing on the system output, the original long SUT waveform can be reconstructed through waveform stitching.

**Figure 2 | Schematic and performance of the low-aberration time magnifier. a,** Experimental setup of the low-aberration time magnifier with the parametric time lens implemented through four-wave mixing (FWM) in a 50-m highly-nonlinear fibre (HNLF). To minimize the third-order-dispersion-induced aberration, both the input and the pump dispersions are provided by proper combination of dispersion compensating fibre (DCF) and large effective-area fibre (LEAF). **b,** The optical spectrum after the HNLF when measuring a 470-fs pulse. The narrow-band idler is filtered out, dispersed, and amplified to become the final output of the time magnifier. **c,** The output time (black, left axis) and pulsewidth (blue, right axis) of the time magnifier as a function of the input time, measuring a temporal-magnification ratio of 61.5 and a consistent temporal resolution of 740 fs across the 300-ps recording window. **d,** The normalized output waveforms of the system when the input pulse is temporally shifted across the recording window. The corresponding output waveform at each input time are labelled by a different colour. Inset: the impulse response of the photodetector. MLL, mode-locked laser; EDFA, erbium-doped fibre amplifier; BPF, band-pass filter; PC, polarization controller; WDM, wavelength-division multiplexer.

**Figure 3 | Schematic and performance of the optical buffer. a,** Experimental setup of the optical buffer, which generates multiple high fidelity replicas of arbitrary signals under test (SUTs) with fine-tunable period for subsequent stroboscopic signal acquisition. A SUT will be loaded into the buffer through the 50/50 coupler, and one replica will be generated when the SUT is circulated for each cavity round trip. **b,** The output waveform when an arbitrary SUT is optically buffered for ten times. **c, d** and **e,** Three examples of buffering performance using distinct SUTs. Ten replicas

are overlapping together (grey curve) and compared to their average (blue). **f,** Quantitative characterization of the buffering performance using 20 different SUTs. Each column set in purple represents the cross-correlation coefficient between the replica after different times of buffering and the original waveform (first replica). The average values among 20 examples are shown by blue triangles, showing a fidelity loss of less than 1% after ten times of buffering. AM, amplitude modulator; EDF, erbium-doped fibre; PC, polarization controller; WDM, wavelength-division multiplexer.

**Figure 4 | Implementation of stroboscopic signal acquisition. a,** Simplified experimental setup of the PARTI system, emphasizing key components for electronic synchronization. A repetition-rate-stabilized mode-locked laser (MLL) and a photodetector (PD) together generate the clock signal for the whole system. An arbitrary waveform generator (AWG) and a delay generator (DG) provide the electrical driving patterns based on the clock signal. Finally the three amplitude modulators (AMs) convert the electrical driving patterns to the optical domain, which control the signal-under-test (SUT) loading (AM1), optical-buffer switching (AM2) and time-magnifier-pump generation (AM3) respectively. **b,** Detailed timing chart of the system with a frame rate of 2 MHz. The driving patterns for AM1, AM2 and AM3 as well as the generated replicas are shown schematically in red, purple, blue and orange, respectively. The corresponding pulsewidth and periods are also labelled on the figure. The vertical black dashed line separates the two consecutive frames. **c,** The experimental electrical driving patterns for the three AMs, using same colours as **b**. In practice, AM2 is only opened for 9 times in each frame, since half of the original SUT that directly passes the buffer without being circulated is also considered as one of the 10 replicas. **d,** Optical waveform (black trace) of pump and SUT combined together when the system is operated according to the timing chart in **b**. The pumps scan through the SUT replicas at a step of 150 ps, thus realizing the stroboscopic signal acquisition. Inset, Zoom-in waveform at the beginning and the end of the scanning. SUT is painted as orange to be clearly differentiated from the blue pump.

**Figure 5 | Dissipative Kerr soliton dynamics measured by the PARTI system. a,** An example 2D evolution portrait, depicting soliton fusion dynamics and transition from a triplet soliton state to a singlet soliton state. **b**, **c**, and **d**, Measured waveforms at different evolution time slices in each stage, illustrating stable triplet solitons at the beginning stage (**b**), soliton fusion at the middle stage (**c**), and stable singlet soliton at the final stage (**d**). **e,** Another example 2D evolution portrait, showing an evolution from doublet solitons to triplet solitons and eventually to a singlet soliton. **f, g,** and **h,** Measured waveforms at different evolution time slices in each stage, illustrating soliton repulsion at the beginning stage (**f**), soliton attraction at the middle stage (**g**), and the stable singlet soliton at the final stage (**h**). For **a** and **e**, black dashed curves emphasizing the soliton transition traces are plotted against the 2D portrait. For **b-d** and **f-h**, three waveforms at different time slices

are plotted in each stage, which are represented in black, blue and red, respectively. Coloured arrows indicate the temporal positions where the waveforms with the corresponding colours are measured. Vertical black dashed lines are plotted according to the soliton positions of the first slice to emphasize the position change of solitons at different time slices.

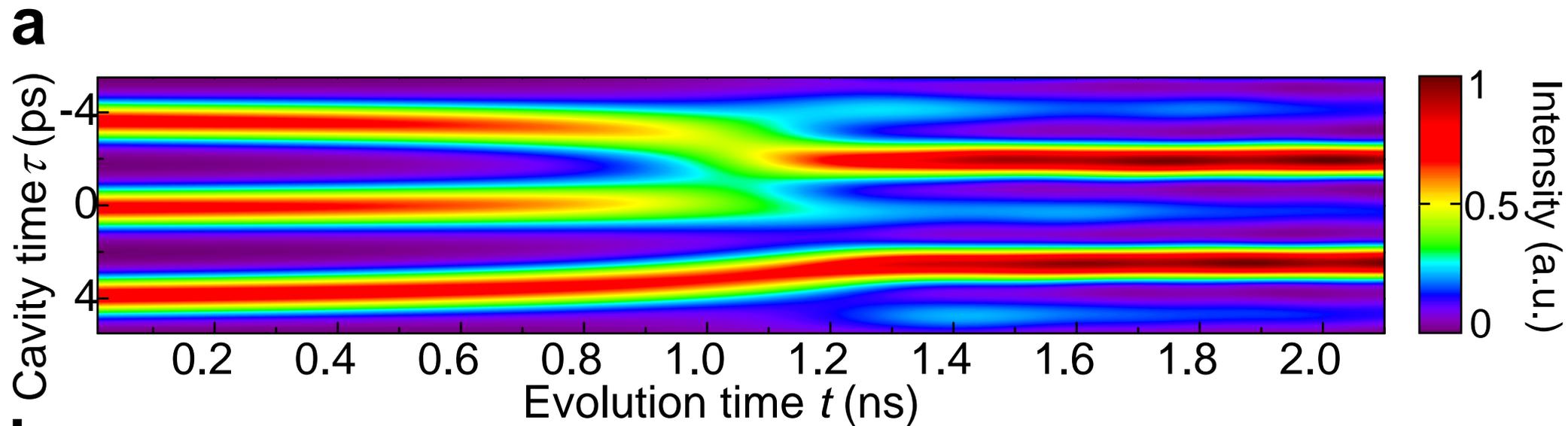

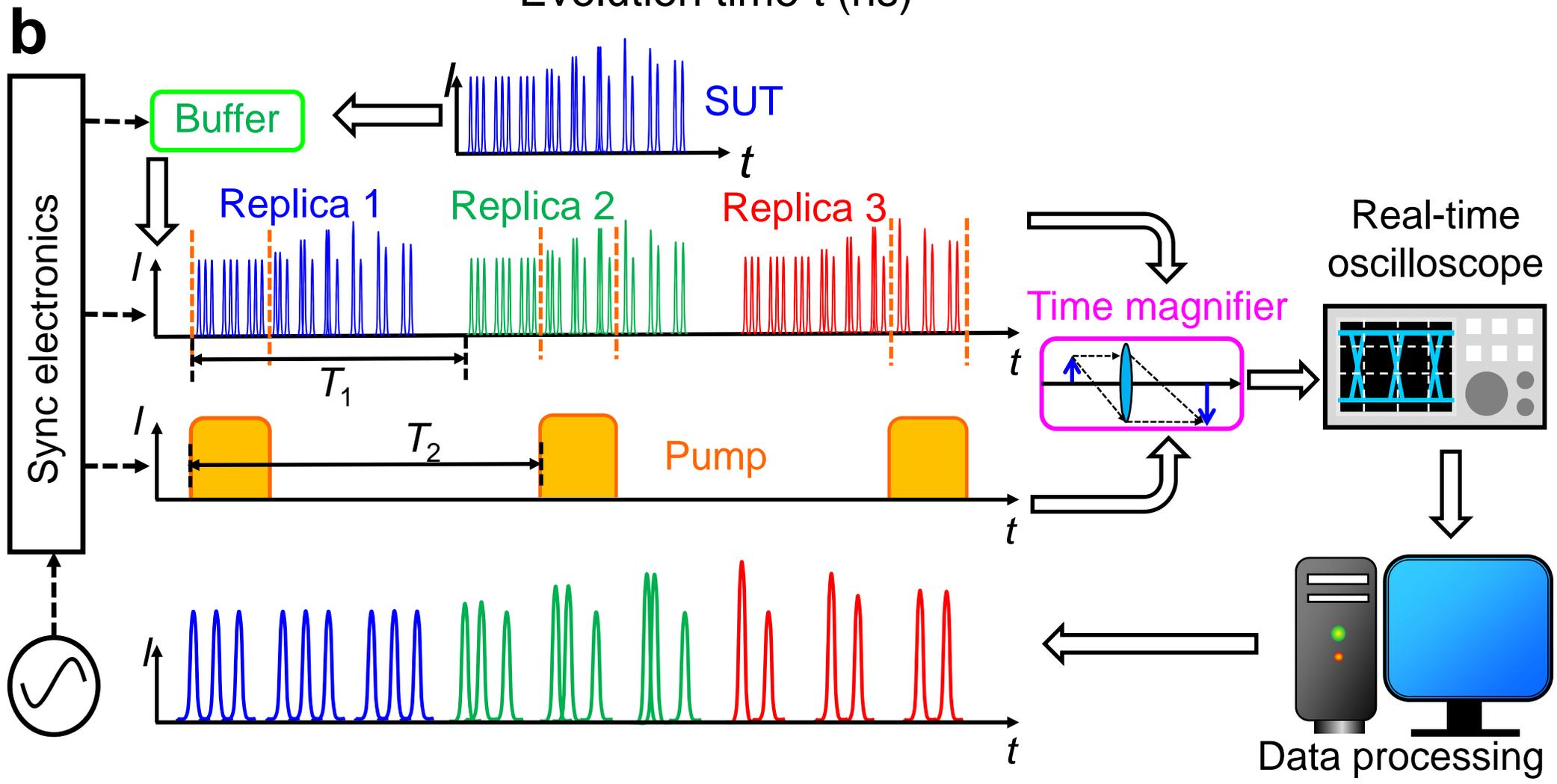

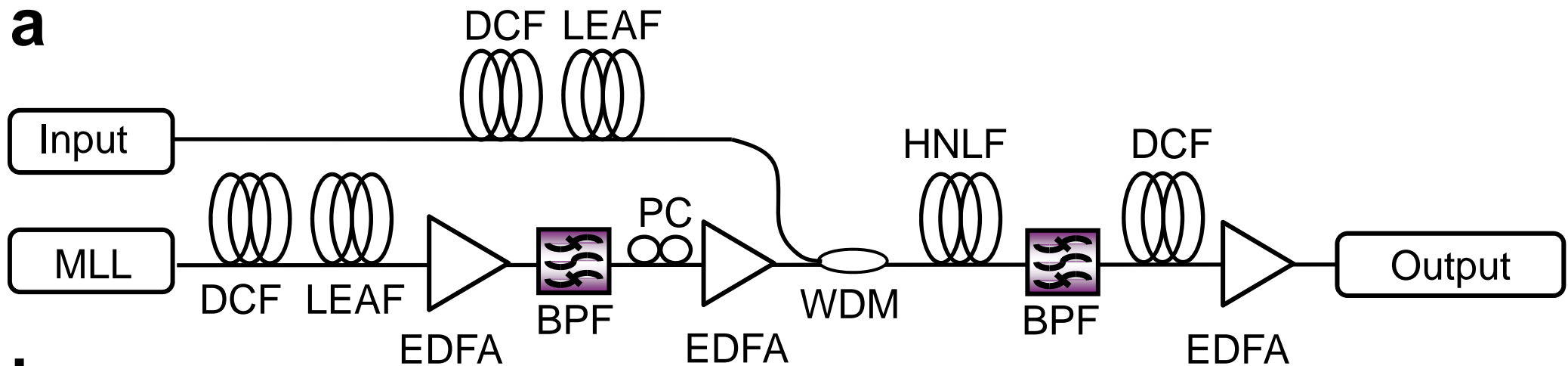
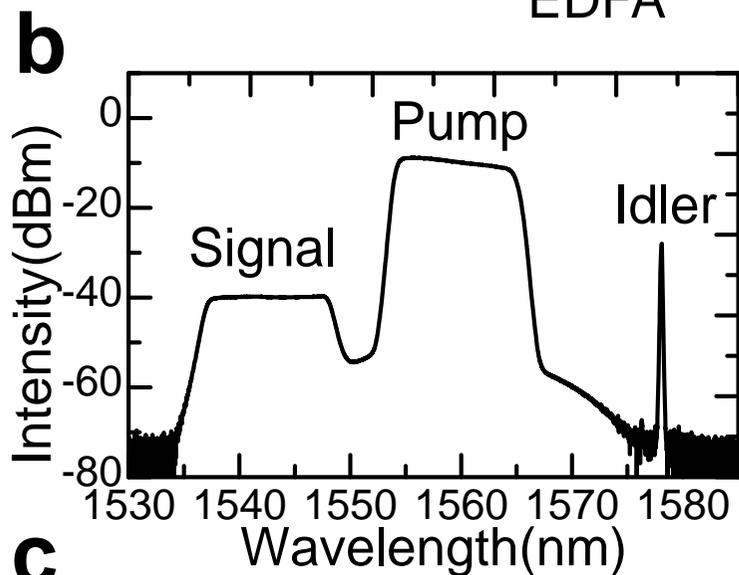
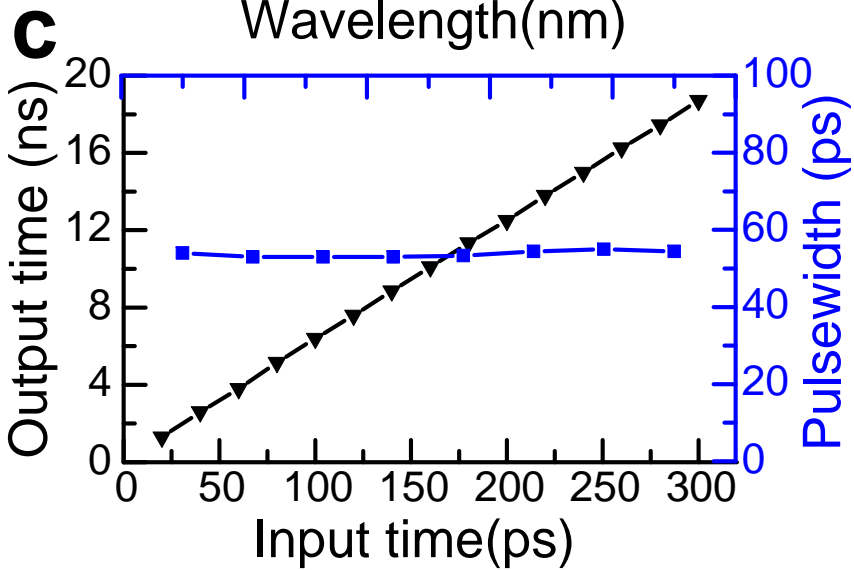
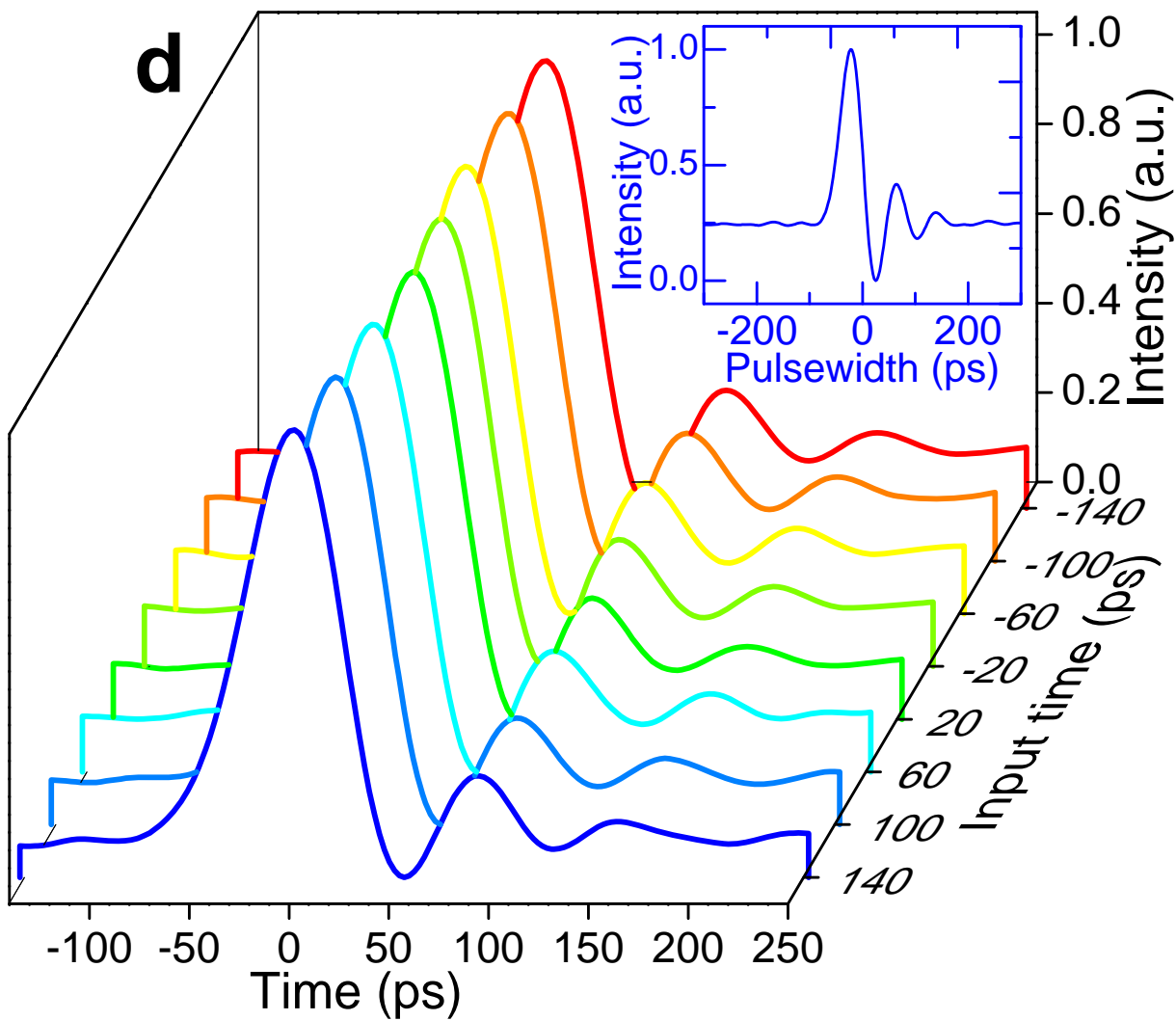

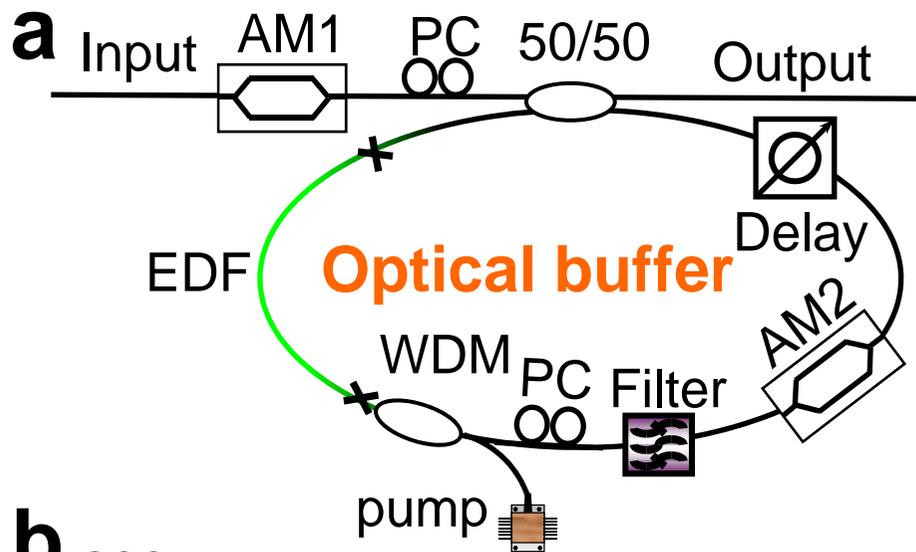
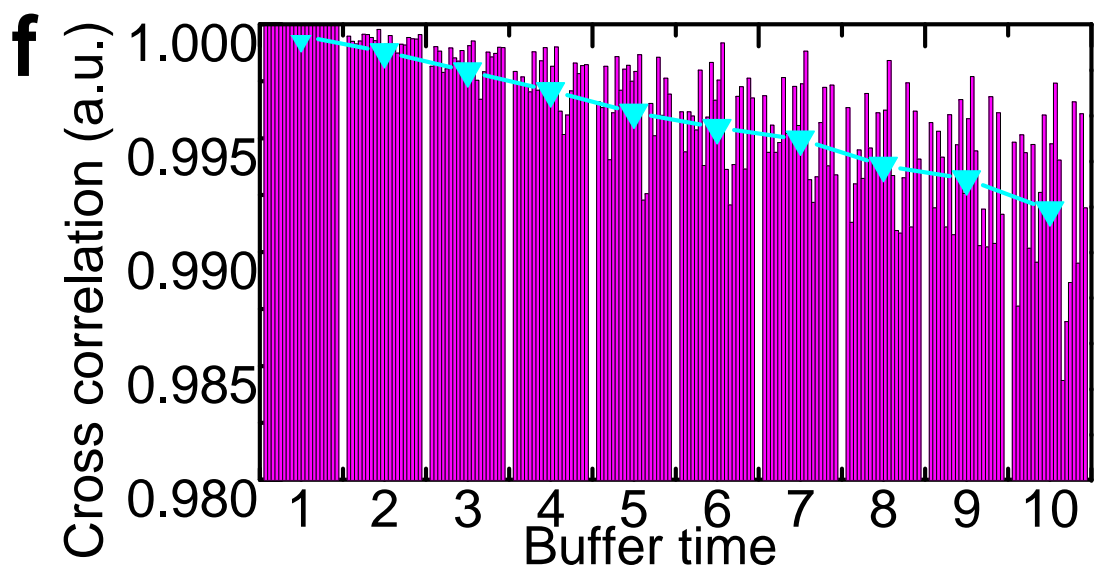
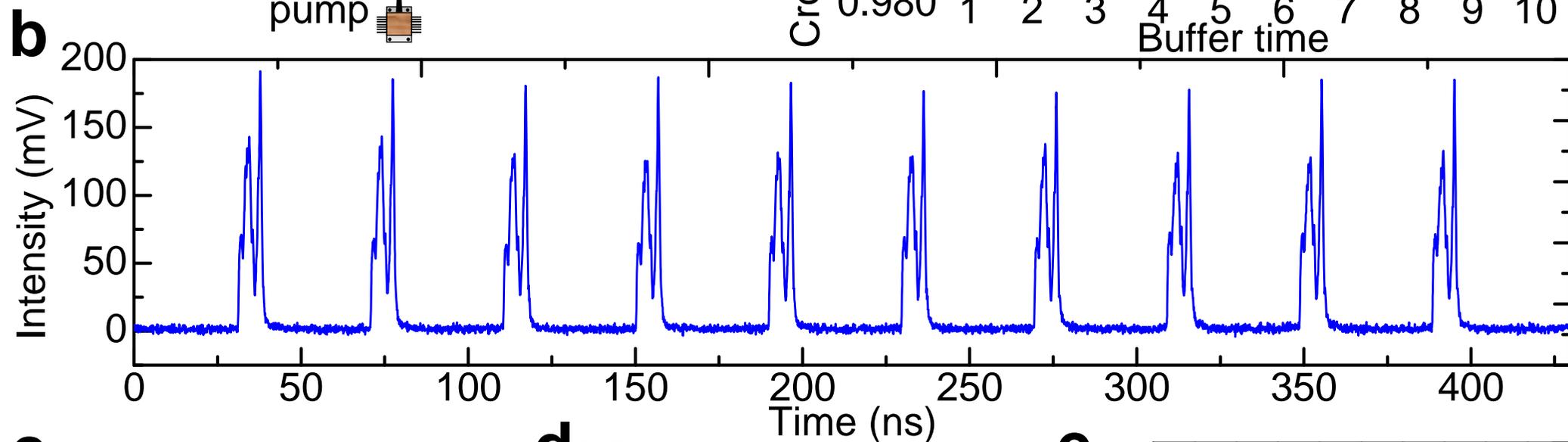
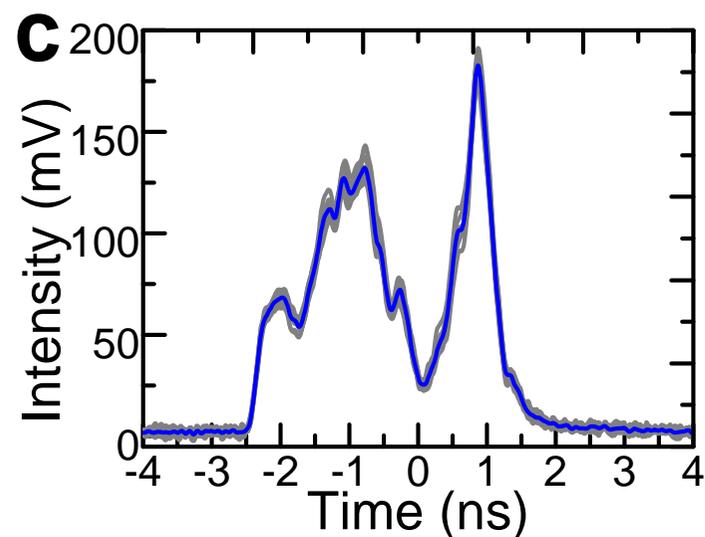
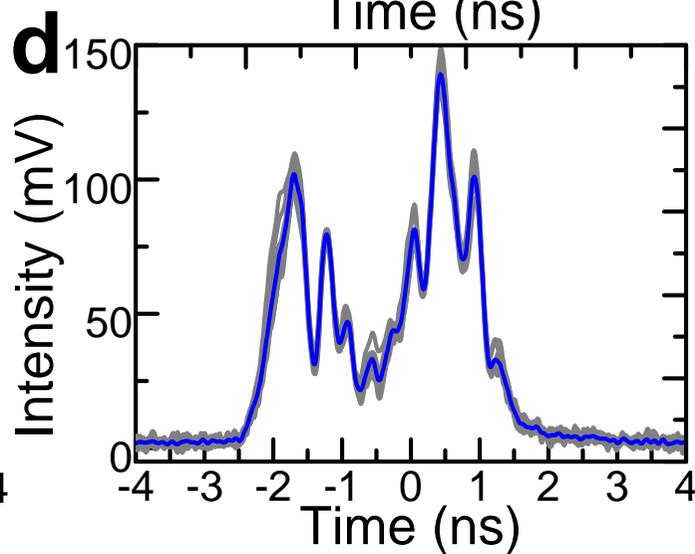
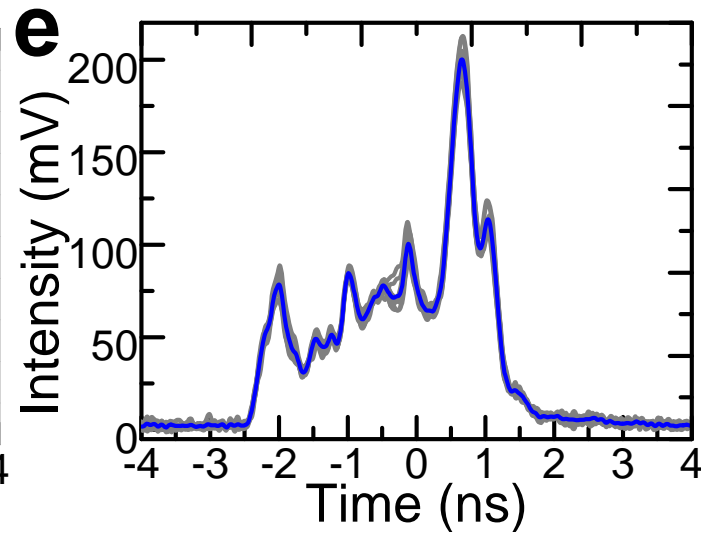

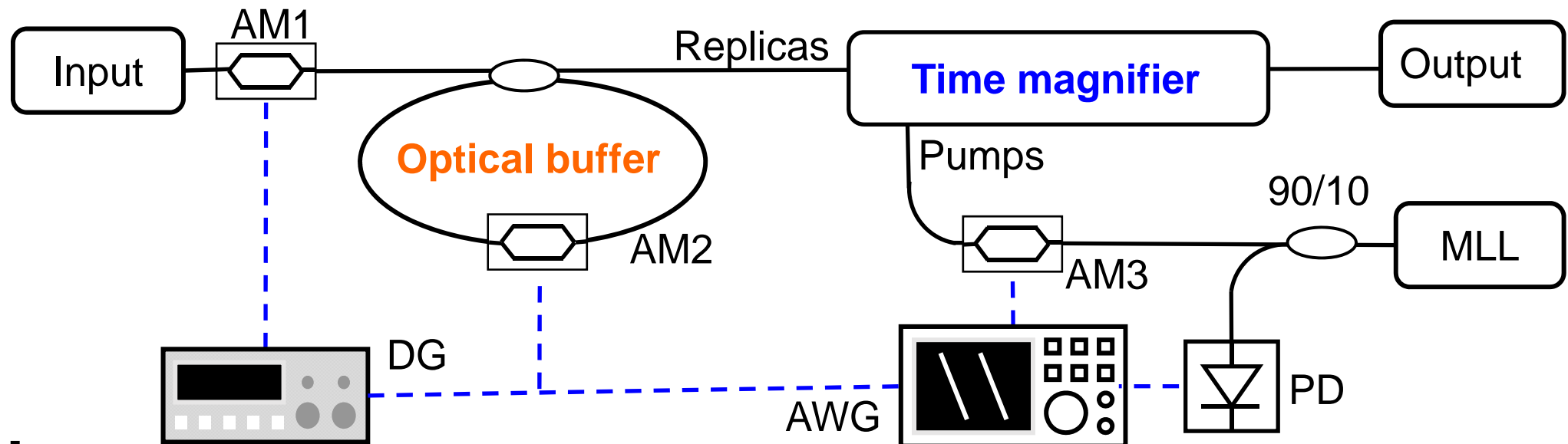
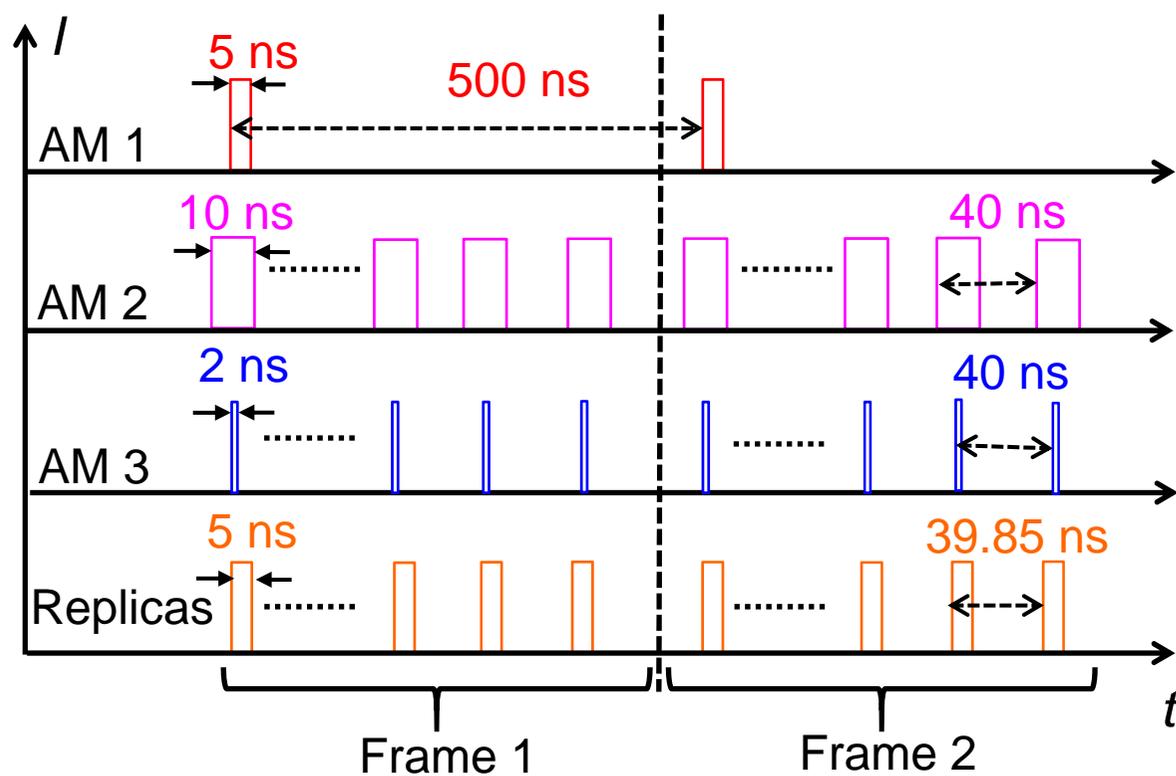
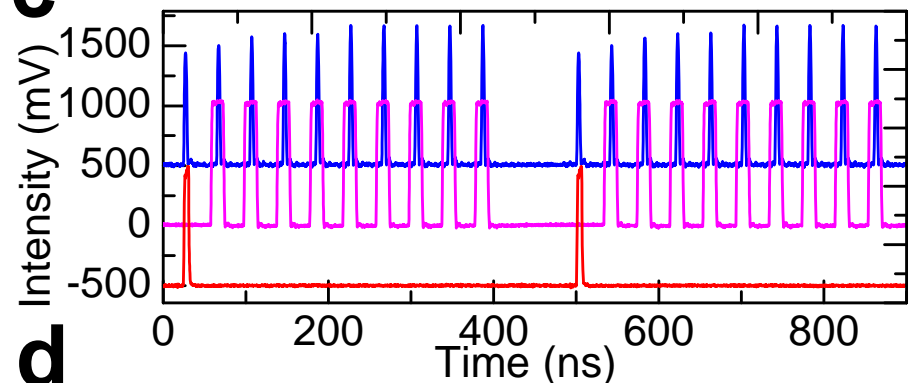
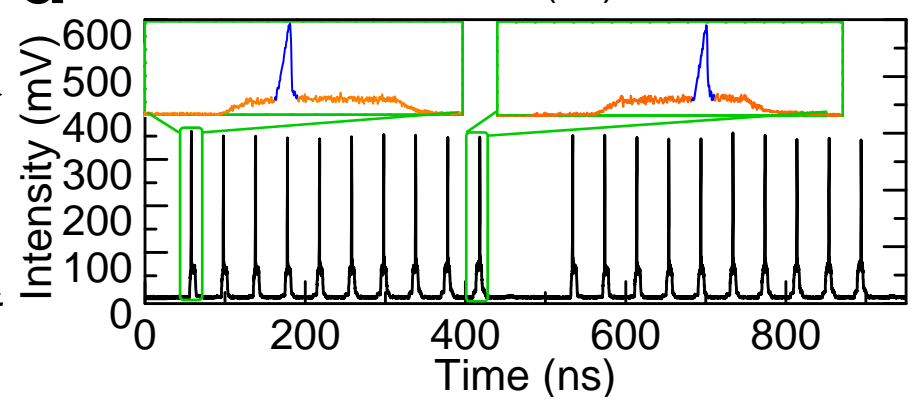

# Supplementary Information

**Supplementary Note 1: Dispersion and spectral intensity response of the buffer**

In this section, the residual dispersion and the spectral intensity response of the optical buffer is characterized, which are responsible for the finite pulse-shape distortion during the buffering. First of all, the residual dispersion is estimated using an intensity auto-correlator and an optical spectrum analyser (OSA). The loop of the optical buffer is opened and a broad-band femtosecond source singly passes the buffer. The corresponding spectral bandwidth and the pulse width are measured to estimate the net dispersion. To measure the dispersion more accurately, the bandpass filter (BPF) is removed first so that a much broader pulse spectrum can pass the buffer, and therefore the pulse width broadening is much more sensitive to the dispersion. As shown in Supplementary Figure 1a, the original pulse had a 3-dB spectral bandwidth of 63 nm. The corresponding transform-limited pulse width is estimated to be about 56 fs based on a Gaussian-shape approximation. After singly passing the optical buffer, the spectral bandwidth is narrowed down to about 37 nm owing to the limited gain bandwidth of erbium-doped fibre (EDF). Similarly, the transform-limited pulse width is approximately 95 fs. The actual pulse width corresponding to the two spectra in Supplementary Figure 1a, *i.e.* before and after passing the optical buffer is measured to be 131 fs and 148 fs, respectively. Based on the above values, the dispersion is estimated using the following equation [1].

$$\Delta t_{out} = \frac{\sqrt{\Delta t^4 + 16(\ln 2)^2 \Phi_2^{''2}}}{\Delta t} \qquad (1)$$

where $\Delta t$ refers to the original pulse width, while $\Delta t_{out}$ is the pulse width after passing through a group-delay dispersion (GDD) $\Phi_2^{''}$. Since the input pulse is not transform-limited, the initial chirp on the pulse should also be taken into consideration. The largest possible dispersion inside the buffer corresponds to the situation where the initial chirp and the chirp after the buffer have opposite sign. In this case, the GDD of the buffer is calculated to be

$$\Phi_2^{''}\Big|_{\Delta t=56\,\text{fs},\Delta t_{\text{out}}=131\,\text{fs}} + \Phi_2^{''}\Big|_{\Delta t=95\,\text{fs},\Delta t_{\text{out}}=148\,\text{fs}} = 6.12\times 10^{-3}\,\text{ps}^2 \quad (2)$$

which corresponds to the dispersion of about 0.28-m single-mode fibre (SMF). The BPF that is previously taken out of the cavity had a foot print of about 3 cm and the corresponding dispersion is negligible. Under this condition (largest possible dispersion), a transform limited 740-fs pulse, which equals the resolution of the time magnifier, will only be broadened by less than 5% after circulating inside the buffer for ten roundtrips. Therefore, a conclusion can be safely drawn that the influence of net residual dispersion on the fidelity of buffering is negligible.

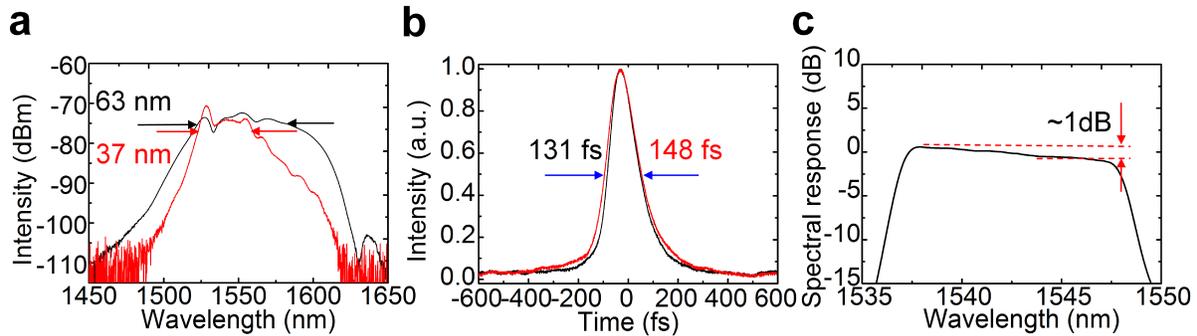

**Supplementary Figure 1 | Characterization of the buffer dispersion and spectral intensity response. a,** Optical spectra of test pulse before (black) and after (red) single passing the buffer. **b,** Pulse width before (black) and after (red) the buffer. **c,** Spectral response of the buffer.

In addition to the net residual dispersion, the non-ideal spectral intensity response (or gain spectrum) of the buffer may also induce pulse shape distortion. In Supplementary Figure 1c, the normalized overall spectral intensity response is shown, which is obtained by multiplying the gain spectrum of EDF and the spectral response of the BPF. As shown in the figure, the actual response is slightly tilted and the value at short wavelength side is about 1 dB higher than that at the long wavelength side. The non-uniformity of the EDF gain and the BPF contribute almost equally to the tilted response. As a consequence, the spectral shape of the original waveform will be slightly narrowed after multiple times of buffering. However, as has been shown in the main text Figure 3, for an un-chirped waveform, this tilted spectral intensity response together with distortions from dispersion and amplification noise only cause less than 1% deviation from original envelope after ten circulations. Therefore the current performance of the optical buffer is still acceptable for proof-of-concept demonstration. Ultimately, the optical gain equalizer can be incorporated to the cavity to further improve the performance, if necessary.

**Supplementary Note 2: Generation of arbitrary waveform**

The arbitrary waveforms in the main text are generated from a continuous-wave (CW) pumped ultrahigh-$Q$ microresonator. As shown in Supplementary Figure 2a, the optical spectrum consists of multiple equidistant spectral lines with a spacing about 89 GHz, which has been band-pass filtered to fit the measurement range of the time magnifier. Since the spectral lines are not phase-locked to each other, the spectral phase changes continuously in spite of the stable spectral shape. Therefore, the corresponding waveform is also evolving all the time, which provides the non-repetitive arbitrary waveform as signal-under-test (SUT). The twenty groups of buffered SUTs used to characterize buffering performance in main text Figure 3f are shown in Supplementary Figure 2b. The buffer loads a section of arbitrary waveform from the ultrahigh-$Q$ microresonator around every 500 ns and generates ten high-fidelity replicas for each loaded SUT. In Supplementary Figure 2b, the output waveform from the buffer is recorded for about 10 μs and therefore includes 20 consecutive buffered groups, where a red dashed circle indicates one group. As observed, the buffered waveforms had distinctively different intensities, while in each group, the ten replicas show similar height with finite fluctuation induced by noise. As an example, the zoom-in of the 12[th] group is shown in Supplementary Figure 2c (same figure as main text Figure 3b).

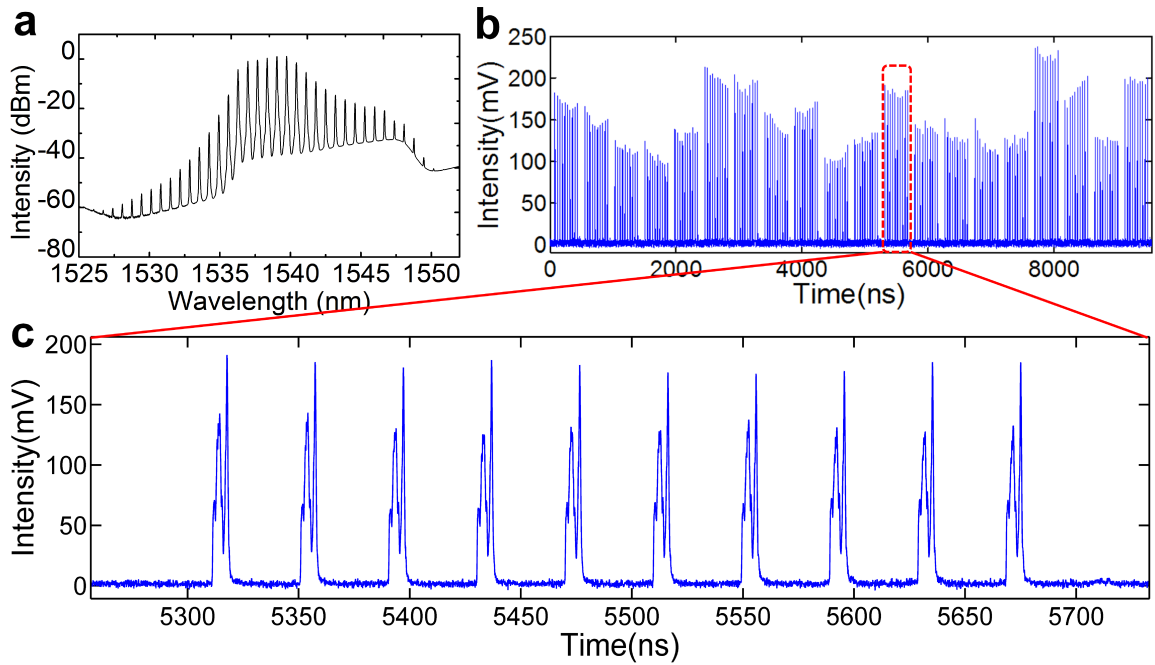

**Supplementary Figure 2 | a,** Optical spectrum of the constantly-evolving arbitrary waveform. **b,** Waveform showing 20 groups of buffered SUT and each group is separated by 500 ns. **c,** Zoom-in of the 12[th] group in **b**.

**Supplementary Note 3: Waveform stitching process**

**Intensity calibration**

In order to reconstruct a long dynamic waveform from ten mosaic magnified sections, data post-processing is required to calibrate the intensity of each frame of magnified waveform before waveform-stitching can be conducted. Each time a section of SUT is loaded into the system, the PARTI system will output ten sections of magnified waveforms, which corresponds to ten consecutive positions on the SUT, as shown in Supplementary Figure 3b. However, as can be observed on each of the ten frames, the magnified waveform sections show a similar envelope shape even though different SUT positions are measured. This can be understood by looking at the four-wave mixing (FWM) spectrum of the time lens measured after the highly-nonlinear fibre (HNLF), as shown in Supplementary Figure 3a. Spectral components from 1538 nm to 1547 nm are filtered out and launched into the parametric time lens to mix with the swept pump centred at 1560. The idler centres around 1580 is generated, which is filtered out and then goes through output dispersion to generate waveforms in Supplementary Figure 3b. Owing to the non-flat spectrum of the swept pump, as well as the different phase-matching condition of the FWM at different pumping wavelength, the parametric conversion efficiency of the time lens at different input time is actually different, which gives rise to the tilted envelope in each output frame. Intuitively, this is equivalent to a spatial thin lens that has different transmission coefficients at different positions. This will degrade the waveform stitching quality, since the magnified waveforms that correspond to the same position on the SUT may show different intensities in two consecutive output frames, which results in discontinuity at the stitching area. Consequently, intensity calibration according to the time lens responsivity curve is required.

To obtain the responsivity curve of the time lens, we add up 2,500 output frames when the dynamic changing waveform is continuously measured by the PARTI system and the result is shown in Supplementary Figure 3c. Since 250 totally different sections of SUT have been measured, the envelope shown in Supplementary Figure 3c should be safely attributed to the responsivity of the time lens. After performing further digital smoothing and normalization, the responsivity curve is obtained, as shown in Supplementary Figure 3d. With the calibration curve, we are able to perform intensity calibration to individual output frames from the system. Supplementary Figure 3e shows the zoom-in view of one frame in Supplementary Figure 3b, and the overall envelope matches well with the responsivity curve we obtain. After the calibration, the reshaped waveform is shown in Supplementary Figure 3f, which shows much better intensity uniformity and can thus be further used for waveform stitching.

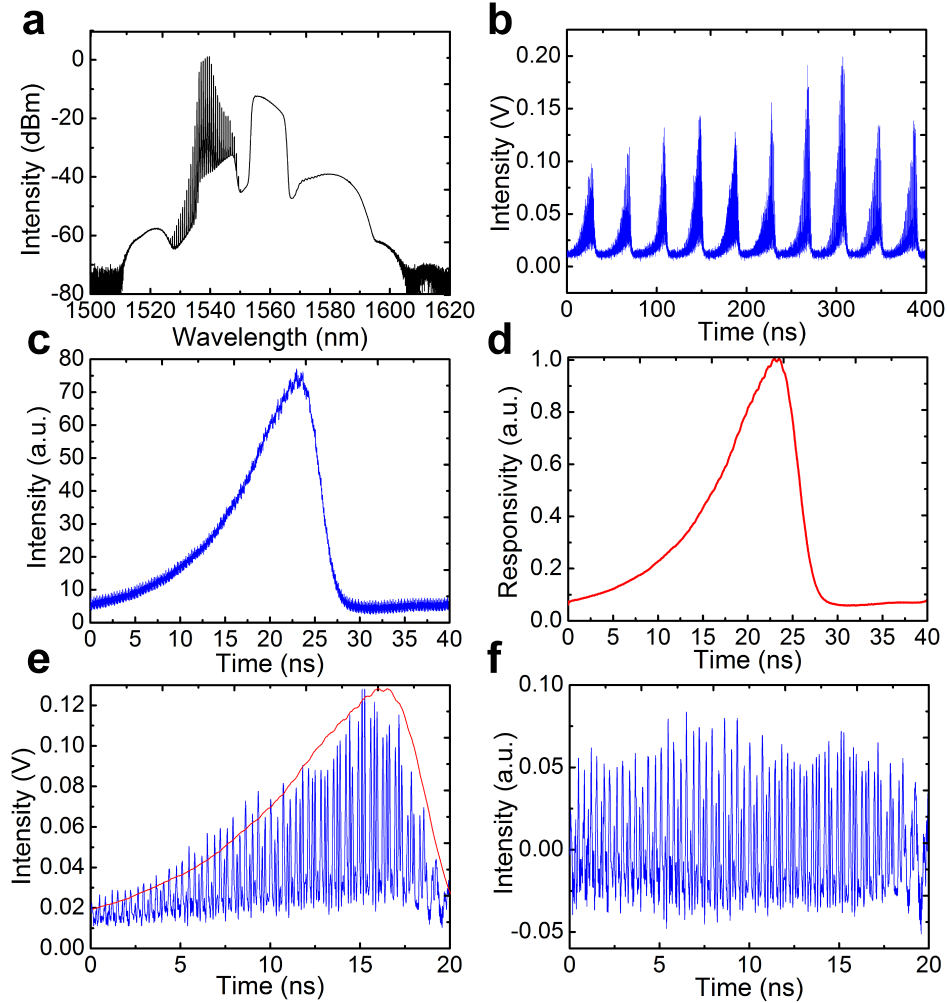

**Supplementary Figure 3 | Measurement and calibration of dynamic waveform. a,** four-wave mixing (FWM) spectrum for the dynamic waveform measurement after the highly-nonlinear fiber (HNLF). **b,** Output waveform from the panoramic-reconstruction temporal imaging (PARTI) system. For each buffered signal-under-test (SUT), system will output 10 frames of waveform, which correspond to 10 consecutive positions on the buffered SUT. **c,** The waveform after adding up 2500 output frames. **d,** Smoothed and normalized curve from **c**, which represents the responsivity across the input time-window and is used to calibrate the intensity of output waveforms. **e,** The zoom-in of the fifth frame in **b** (blue curve) together with the calibration curve (red curve). **f,** The waveform after intensity calibration.

**Waveform stitching**

Supplementary Figure 4 describes the process for waveform stitching, which consists of three main steps: initial temporal alignment, stitching-area identification and fine temporal alignment. In step 1, two consecutive frames of magnified waveforms (blue and red curves) that correspond to the evolution time 0.73 ns~1 ns in main text Figure 5e (case 2) are shown in Supplementary Figure 4a as an example. Each section contains 2000 data sampling points, corresponding to 20 ns on the magnified waveform and 320 ps on the original SUT. The initial alignment is made according to the predefined scanning step size in stroboscopic acquisition, i.e. 150 ps in this work. Considering the temporal magnification ratio of 61.5, a step size of 150 ps on the SUT corresponds to about 9.23 ns, or 923 data points on the magnified output waveform. Therefore, the blue trace and red trace are relatively shifted accordingly as initial temporal alignment. Vertical offset is used for easier waveform comparison. As observed in the zoom-in view of the temporally overlapping area (1077 data points) of the two traces in Supplementary Figure 4b, the repetitive three-pulse structure in two traces already matches well with each other, which in turn confirms the accuracy of the initial alignment. Ideally, the overlapping areas in both traces should appear identical, such that any point within the area can serve as the stitching point between the two traces. However, because of the non-uniform responsivity of the time lens, the trailing area of the blue trace as well as the leading area of the red trace show larger distortion on the waveform owing to poorer signal-to-noise ratio (SNR). Consequently, an optimal stitching area exists where both traces have reasonable SNR and thus exhibit highest resemblance. Therefore, in step 2, the optimal stitching area is identified by calculating the cross-correlation coefficient between the two traces. Note that the cavity roundtrip time of the microresonator is about 11.3 ps, which corresponds to about 70 data points on the magnified waveform. Therefore, the overlapping areas in both traces can be segmented into 15 sections according to the roundtrip time of the microresonator, which function as 15 comparison pairs. The maximum values of the cross-correlation function between each pair of waveform sections are calculated to identify the optimal stitching area as shown in Supplementary Figure 4c. During the calculation, a comparison window is slightly expanded to 80 data points such that slight temporal misalignment between the two traces will not affect the maximum cross-correlation coefficient. As observed in Supplementary Figure 4c, the coefficient first rises and then falls, which matches well with the SNR evolution. The maximum value is located at $10^{th}$ roundtrip and it is therefore selected as the stitching area, denoted by the green dashed region in Supplementary Figure 4b. In the final step, fine temporal alignment is achieved according to the cross-correlation function to locate the precise stitching point. The cross-correlation function between the $10^{th}$ waveform sections in both traces is shown in Supplementary Figure 4d. The maximum correlation value is reached when the two waveforms are relatively shifted for one data point (labelled by green triangle), which indicates that the red trace should be shifted by one data point for fine temporal alignment during the data stitching. To avoid splitting the single roundtrip waveform during the stitching, the starting point of the 10$^{th}$ roundtrip waveform is used as the stitching point, and the data from the two traces are combined accordingly, which generates the stitched waveform shown in Supplementary Figure 4e. Since the SNR evolution is similar for the overlapping areas between any two neighbouring frames, the approach above is adapted throughout the rest of the data processing. Just like the process demonstrated above, for each buffered SUT, the ten magnified waveforms can be stitched together to reconstruct a 1.5-ns long waveform (de-magnified time scale), which depicts the dynamic dissipative-soliton-evolution process with a temporal resolution of 740 fs, representing a time-bandwidth product about 5 times larger than the previous record value demonstrated in conventional temporal imaging systems. To visualize the roundtrip waveform evolution process, the 1.5-ns long waveform is sectioned according to the roundtrip time, i.e. 11.3 ps, and the 2D evolution map shown in Figure 5 in the main text can be obtained.

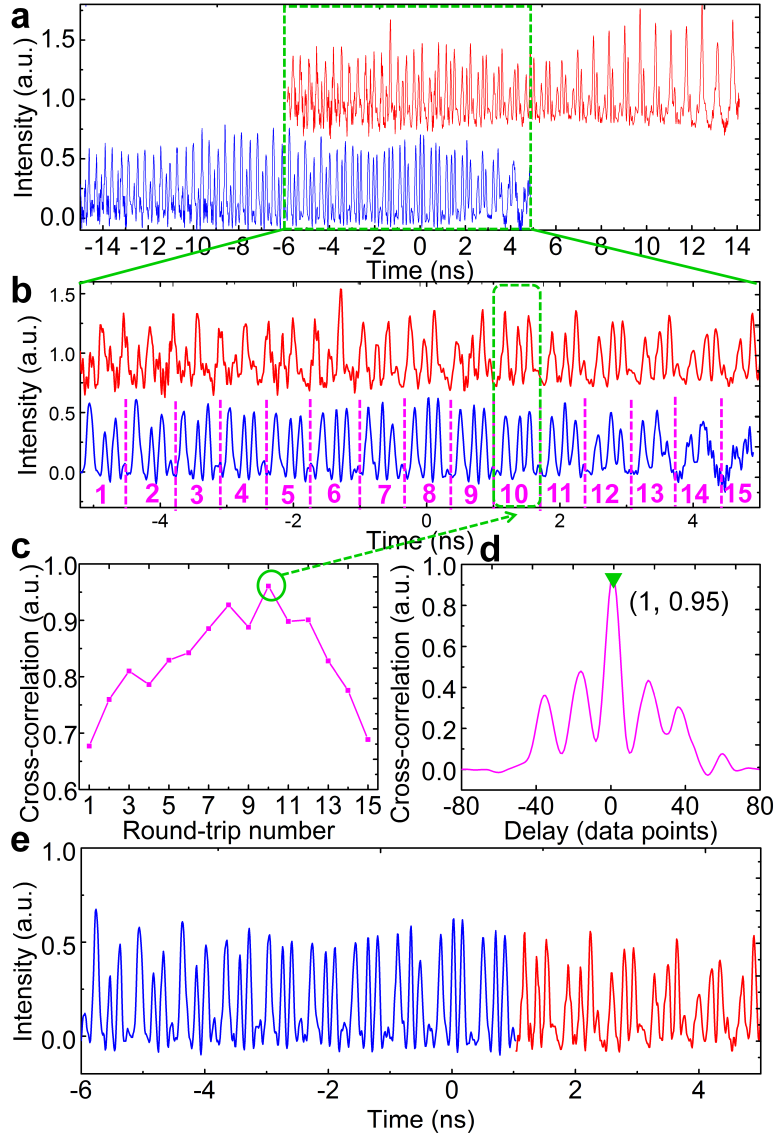

**Supplementary Figure 4 | Waveform stitching process. a,** The magnified waveform in two consecutive output frames (blue, red) are temporally aligned according to the scanning step size in stroboscopic acquisition and are vertically offset for comparison. **b,** Zoom-in of the temporal overlapping area in **a**. Each trace is segmented into 15 sections according to the roundtrip time of the microresonator, which serve as 15 comparison pairs. **c,** The maximum values of cross-correlation coefficients between each comparison pair in the blue and red traces. The maximum value labelled by the green circle indicates the optimal stitching area, while the lower values are

attributed to the low SNR in either red or blue trace. **d,** The cross-correlation function between the 10th waveform sections in blue and red traces. The horizontal location of the peak indicates the amount of shift required for fine temporal alignment. **e,** Longer waveform after stitching according to the identified stitching point.

**Supplementary Note 4: Estimation of scalability.**

Owing to the requirement for repetitive amplification to overcome the loss inside the optical buffer, the amplified spontaneous emission (ASE) noise will accumulate, which degrades the optical SNR of the signal. As this effect is universal for all amplifiers, the SNR degradation is unavoidable, which will limit the ultimate scalability of the PARTI system. To find out the maximum allowable buffering number of times and thus the ultimate performance of the PARTI system, the noise accumulation inside the buffer is theoretically analysed.

The repetitive amplification inside an optical buffer is essentially equivalent to a long-haul transmission system, where erbium-doped fibre amplifiers (EDFAs) are deployed periodically to compensate the transmission loss. In fact, circulating loop transmission experiments have been conducted to study the long-haul transmission system [2]. Typically in the long-haul transmission system, the EDFA gain is just enough to compensate the fibre loss in each section such that the signal intensity remains relatively constant, which matches exactly with our case inside the optical buffer. Therefore, the total ASE power $P_{sp}$ after N times of buffering can be expressed by [3]:

$$P_{sp} = 2n_{sp} h v_0 N(G-1) \Delta v_{opt} \tag{3}$$

where $h$ is the Planck's constant, $\nu_0$ is the central frequency of the buffered signal, $N$ is the buffering time, $G$ is the gain of the intra-buffer amplifier and $\Delta\nu_{opt}$ is the bandwidth of the optical filter after the amplification. $n_{sp}$ is the population inversion factor and can be expressed as $n_{sp} = \dfrac{N_2}{N_2 - N_1}$, where the $N_2$ and $N_1$ are atomic populations for the excited and ground states of erbium, respectively. The precise value of $n_{sp}$ can be found by solving the rate equations [3], which varies along the EDF and depends on the pump and signal power. While it is beyond our scope to find the exact value, it can be conveniently estimated by

$$n_{sp} \approx \frac{1}{2} F_n \qquad (4)$$

where $F_n$ is the noise figure of EDFAs [3]. The state-of-the-art pre-amplifiers can achieve a noise figure of 4.3 dB (e.g. Amonics AEDFA-PA-30). Therefore, the value of $n_{sp}$ is adopted accordingly to be 1.346. In addition, according to the current buffer parameters, $G$ is 15.85 (12dB) and $\Delta\nu_{opt}$ is 1.25 THz (10 nm). $h\nu_0$ is around 0.8 eV at 1.55 μm. Inserting these values into equation (3), it is obtained that the accumulated noise power after 100 times of buffering (N=100) is about 0.64 mW. Since this power level is unlikely to saturate the EDFA, the gain can be regarded as constant for the signal so as to keep the same signal power in each circulation. According to our typical experimental value, the average intra-cavity signal power is 1.25 mW, which is comparable to the accumulated ASE power. However, note that the signal has a temporal duty ratio of around 1/8 (5ns versus 40-ns cavity time), while the ASE power is evenly distributed over time. Consequently, the peak power of signal power is 10 mW and therefore the optical SNR after 100-time buffering is about 12 dB based on the theoretical analysis shown above. Meanwhile, the optical SNR is not the only parameter determining the final signal quality obtained from the oscilloscope. According to [3], the electrical SNR is related to the optical SNR as

$$SNR_{el} \approx \frac{\Delta \nu_{opt}}{2\Delta f} SNR_{opt} \tag{5}$$

if only the dominating signal-spontaneous beat noise is considered, where $\Delta f$ is the detection bandwidth of the photodetector. To simplify the analysis, we regard the temporal magnification system and the photodetector as a single unit, which achieves a detection bandwidth of around 500 GHz. The small optical SNR degradation inside the temporal imaging system can be neglected compared to that induced by the optical buffering, since only one stage of parametric conversion and a single low-noise pre-amplifier are involved. Under this circumstance, the electrical SNR of the final signal is estimated to be around 13 dB, slightly lower than the standard criteria (15.6 dB) in optical communication, which guarantees a bit-error rate (BER) of $10^{-9}$ [3]. According to this criterion, the maximum allowable buffering number of times is calculated to be 53. On the other hand, it is also worth noting that the PARTI system operates with high-speed real-time oscilloscopes. For a 20-GHz real-time oscilloscope (same as the one used in the experiment), the effective number of bits (ENOB) is typically about 6 [4], which set the effective dynamic range of signal acquisition at 15 dB. Therefore, specifically for the PARTI system, an electrical SNR of 15 dB is sufficient. Under these circumstances, the maximum allowable buffering number of times is 61.

To realize the 100 times of buffering, the buffer requires further improvement in terms of cavity loss. Currently the intra-cavity variable optical delay line is constructed with a circulator, a fibre collimator, and a mirror mounted on a translation stage, which together induce an attenuation of about 3.5 dB. This is required for finding the optimal operating condition, e. g. the temporal scanning step size. However, when the optimal cavity length is fixed, the variable delay line can be readily replaced by an SMF. In addition, the intra-cavity amplitude modulator (AM) can be replaced by state-of-the–art high-speed optical switches (http://eospace.com/switches.htm) that offers insertion loss of less than 3 dB. Therefore, it is expected that the cavity loss of the optical buffer can be further reduced to around 6 dB, which further reduces the ASE noise by lowering the required gain. Under this circumstance, the SNR is calculated to be around 20 dB after 100 times of buffering, which is sufficient to generate high-fidelity results.

Our simplified estimation matches with the comprehensive derivation in [5], where it is shown that the SNR can be slightly above 15.6 dB after 100 circulations for an input power of -10 dBm at 160 Gbit·s$^{-1}$ data rate. Moreover, in a pulse replication experiment [6], an optical buffer with a similar configuration has successfully generated more than 1,000 replicas of single pulse with SNR > 20dB, which further convince us the potential scalability of PARTI system to be at least 100 times.

**Supplementary References:**